\documentstyle[prb,aps,multicol,psfig]{revtex}
\begin{document}
\draft
\hyphenation{following}
\newcommand{\Eanis}{$E_{anis}$}
\renewcommand{\k}{{\bf k}}
\newcommand{\R}{{\bf R}}
\renewcommand{\r}{{\bf r}}
\newcommand{\spinup}{\left|\uparrow\right>}
\newcommand{\spindown}{\left|\downarrow\right>}
\newcommand{\be}{\begin{equation}}
\newcommand{\ee}{\end{equation}}
\newcommand{\K}{{\bf K}}
\newcommand{\m}{{\bf m}}
\newcommand{\M}{{\bf M}}
%
%
\title{Magnetocrystalline Anisotropy Energy 
of Transition Metal Thin Films:
A Non-perturbative Theory}
\author{A. Lessard, T. H. Moos, and W. H\"ubner}
\address{Institute for Theoretical Physics, Freie Universit\"at 
Berlin, Arnimallee 14, D-14195 Berlin, Germany}
\date{\today}
\maketitle
\begin{abstract}
The magnetocrystalline anisotropy energy $E_{anis}$ of free-standing 
monolayers and thin films of Fe and Ni is determined using two different
semi-empirical schemes.
Within a  
{\em tight-binding calculation} for the 3$d$ bands alone, we analyze in detail the
relation between bandstructure and $E_{anis}$, treating spin-orbit
coupling (SOC) non-pertubatively.  We find important contributions 
to $E_{anis}$ due to the lifting of band degeneracies near the Fermi level 
by SOC. 
The important role of degeneracies is supported by the 
calculation of the electron temperature dependence of the
magnetocrystalline anisotropy energy, which decreases with the
temperature increasing on a scale of several hundred K. In general, $E_{anis}$ 
scales with the square of the SOC constant $\lambda_{so}$.
Including 4$s$ bands and $s$-$d$ hybridization,
the {\em combined interpolation scheme} yields anisotropy energies that 
quantitatively agree well with experiments for Fe and Ni monolayers on
Cu(001).
Finally, the anisotropy energy is calculated for systems of up to 14
layers.  Even after including {\em s}-bands and for multilayers, the importance of degeneracies persists.
Considering a fixed fct-Fe structure, we find a reorientation of the magnetization from
perpendicular to in-plane at about 4 layers.  For Ni, we find the correct
in-plane easy-axis for the monolayer.  However, since the anisotropy
energy remains nearly constant, we do not find
the experimentally observed reorientation.

\end{abstract}

\begin{multicols}{2}
\section{Introduction}
The dependence of the total energy of a ferromagnetic crystal on the
direction of magnetization originates from the magnetic dipole-dipole
interaction as well as from spin-orbit coupling (SOC), as proposed by
van Vleck~\cite{Vleck37}. The magnetic anisotropy energy is expected
to be enlarged in systems of low symmetry, i.e. at surfaces,
interfaces, and thin films~\cite{Neel54} or in one-dimensional systems
such as quantum corrals~\cite{Rock}.  Recently, a magnetization
easy-axis perpendicular to the film plane has been observed  for a wide variety
of thin film systems, for example for thin films of fcc Fe on
Cu(001)~\cite{expts,FB53,Li}.
Some of these systems are promising candidates for
magnetic high-density storage media.

In spite of many theoretical
attempts~\cite{Bennett71,Takayama76,Gay86,Bruno89,Pick,wan93a,kel93,Cinal94,Weinberger},
the relation between the electronic structure and magnetocrystalline
anisotropy energy $E_{anis}$
could not be fully clarified so far.
Some very important questions are subject to intense discussion:
(i) Which bandstructure details lead to significant contributions to 
$E_{anis}$? Especially the treatment of 
degenerate bands near the Fermi level 
has brought up controversies~\cite{Pick,wan93a,kel93}. 
(ii) How does $E_{anis}$ depend on the SOC strength $\lambda_{so}$?
(iii) How is it influenced by the substrate lattice constant?
Moreover, there is no unified thermodynamic and electronic theory to
determine the temperature dependence of $E_{anis}$. Finally, the
correct prediction of magnetic anisotropy for real
systems still remains a challenge, since due to the quenching of
orbital angular momentum in 3$d$ transition metal systems,
$E_{anis}$ is several orders of magnitude smaller than other
contributions to the total energy of a crystal (typically about
$0.1-1$~meV per atom in ultrathin films). 

The magnetic anisotropy of thin films has been investigated using 
two essentially different approaches. In semiempirical
calculations~\cite{Bennett71,Takayama76,Bruno89,Cinal94}, 
the magnetocrystalline anisotropy energy $E_{anis}$
is determined by means of parametrized tight-binding bandstructures. 
Usually, spin-orbit coupling is restricted to second order perturbation
theory. 
On the other hand, {\em ab-initio}\/ calculations have been
made~\cite{Gay86,wan93a,kel93,Weinberger} and lead to realistic
bandstructures.   All calculations make use of the controversial force
theorem~\cite{force}.  Convergence, however, is difficult to achieve;
sometimes, additional assumptions are made in order to obtain
converged results ({\em state tracking
  method}\/~\cite{wan93a}).  

The structure of thin Fe films deposited on Cu(001) has been widely
investigated, especially the dependence of the structure and
magnetization orientation on the temperature.  For films of less than
5 monolayers (ML) deposited at low temperatures, a distorted
fcc-structure is found, with magnetization perpendicular to the film
plane.  At 5 ML, a transition to in-plane magnetization is
observed, as well as a restructuration of the film. 
It is still not clear if this reorientation
transition is an effect of the structural changes taking place in the
film at 4-5 ML~\cite{FB53,FB52,M74,zar,Arvanitis}.

In this paper, we 
investigate a simple quadratic Fe and Ni monolayer and fcc multilayers
systems up to 14 ML epitaxially grown
on the Cu(001) surface and neglect further interactions with the
substrate. The bandstructures are calculated within 
two different semi-empirical schemes, including
SOC completely non-perturbatively
without resorting to degenerate or non-degenerate perturbation theory
of any order.
A {\em tight-binding calculation} of the
3$d$-bands allows 
for a detailed, {\bf k}-space resolved analysis 
of the role of degeneracies for $E_{anis}$.  
It is shown that degeneracies located near the Fermi level can yield
significant contributions, if they occur along {\em lines} in {\bf k}-space.
We find for these that generally $E_{anis}\propto\lambda_{so}^2$ holds.
Including 4$s$-bands by means of the {\em combined interpolation
scheme}~\cite{Hodges66} and fitting the parameters to {\em ab-initio}\/
calculations, we obtain the correct sign and values of $E_{anis}$ for
the systems considered with this fully convergent method. 
That could be achieved neither by a fit using bulk
parameters nor by employing a real-space density of states
calculation, the so-called recursion method~\cite{Pick}.
Moreover, we find the characteristic scale for the temperature
dependence of the magnetic anisotropy to be $\lambda_{so}$, rather than
the bandwidth. This supports the significance of the lifting of
degeneracies at $E_F$ by $\lambda_{so}$ and demonstrates 
the importance of contributions to magnetic anisotropy 
due to Fermi-edge smearing.

Finally, we calculate the anisotropy energy of multilayer systems.
For systems of tetragonally-distorted Fe of 2 to 14 ML, we find a
transition from magnetization perpendicular to the
plane to in-plane magnetization at about 4 ML.  We conclude from our
calculation that the experimentally observed reorientation at 5 layers
is not necessarily caused by a structural phase transition.  For Ni, we
find a nearly constant anisotropy energy
from the fourth layer on, in disagreement with the results of Schulz
and Baberschke~\cite {Schulz94}, who find a reorientation from in-plane  to parallel
magnetization at 7 ML.  In both cases, the degeneracies
near the Fermi-level are found to play an important role for the
dependence of the anisotropy energy on the film thickness.

This paper is organized as follows: In Section II, the
interpolation schemes (II.A, II.B, II.C) and the determination of $E_{anis}$ (II.D) are presented. 
The results for the tight-binding scheme for $d$-bands alone are shown
in section III.A,
the role of degeneracies is analyzed in detail in III.B while the
results for the complete $s$- and $d$-band calculation for Fe and Ni
monolayers on Cu(001) and other substrates are given in section III.C.
The influence of crystal
field splitting is investigated. Some aspects of the temperature
dependence of magnetic anisotropy are considered in III.D, and the
results for multilayer systems are presented in III.E and III.F.
Section IV sums up the most important
results.

\section{Theory}
\subsection{Bandstructures}
The magnetocrystalline anisotropy energy $E_{anis}$ depends sensitively
on the electronic structure of the system. To simplify the analysis, the 
bandstructure of the monolayer is calculated in two steps. First, the
3$d$-bands are described within a tight-binding scheme. Although the
resulting $E_{anis}$ as a function of the 3$d$-bandfilling $n_d$ shows
already the most important features, the 4$s$-bands and 
$s$-$d$-hybridization have to be taken into account for a correct numerical 
evaluation of $E_{anis}$.

For the 3$d$-bands, the tight-binding formalism introduced by
Fletcher~\cite{Fletcher52} and Slater and 
Koster~\cite{Slater54} is 
adapted to the monolayer. 
The Hamiltonian $H^d=H_{at}+\Delta U$ is set up as a $10\times10$
matrix with respect to the basis of Bloch wave functions 
\begin{equation}
\label{dbasis}
\psi_{n{\bf k}}({\bf r})=\frac{1}{\sqrt{N}}\sum_{\bf R}e^{i{\bf
    k}\cdot{\bf R}}\phi_n({\bf r}-{\bf R}).
\end{equation}
Here, $H_{at}$ is the atomic Hamiltonian, $\Delta U$ the additional 
crystal field in the monolayer.
$\phi_i$, $i=1,...,5$ ($i=6,...,10$) 
are the atomic 3$d$ orbitals commonly denoted by $xy$, $yz$, $zx$, 
$x^2-y^2$ and $3z^2-r^2$ respectively, together with the spin
 eigenstate $\spinup$ 
($\spindown$) with respect to the spin quantization axis $z_M$. 
In the simple quadratic monolayer, only orbitals located on neighboring
atoms are included. 
The extension to second nearest neighbors does not lead to 
further insight~\cite{MoosDA}. 
With the $x$- and $y$-axes oriented along axes connecting nearest 
neighbors in the monolayer, the spin-polarized 
Hamilton matrix has (within the three-center approximation) 
the form
\begin{eqnarray}
H^d_{11} &=& E_0+\Delta^V_{\mbox{\footnotesize
    Ni}}+2\tilde{B}_1(\cos2\xi+\cos2\eta) -J_{ex}/2\nonumber\\ 
H^d_{22} &=& E_0+2\tilde{B}_2\cos2\xi+ 2\tilde{B}_3\cos2\eta
 -J'_{ex}/2\nonumber\\
H^d_{33} &=& E_0+2\tilde{B}_3\cos2\xi+ 2\tilde{B}_2\cos2\eta
 -J'_{ex}/2 \nonumber\\
\label{htbfe}
H^d_{44} &=& E_0+\Delta^V_{\mbox{\footnotesize
    Fe}}+2\tilde{B}_4(\cos2\xi+\cos2\eta) 
-J'_{ex}/2 \\ 
H^d_{55} &=&
 E_0+\Delta^V_{\mbox{\footnotesize Ni}}+\Delta^V_{\mbox{\footnotesize
     Fe}}+2\tilde{B}_5(\cos2\xi+\cos2\eta) -J_{ex}/2 \nonumber\\ 
H^d_{45} &=& H^d_{54}= H^d_{9,10}=H^d_{10,9}=
   2\tilde{B}_6(\cos2\xi-\cos2\eta) \nonumber\\
 & \mbox{and} \nonumber \\ 
H^d_{ii} &=& H^d_{i-5,i-5}+J_{ex} \quad \mbox{for } i=6,10, \nonumber \\
H^d_{ii} &=& H^d_{i-5,i-5}+J'_{ex} \quad \mbox{for } i=7,8,9. \nonumber
\end{eqnarray}
Here, $\xi=\frac{1}{2}k_xa$ and $\eta=\frac{1}{2}k_ya$
are the normalized components of the crystal momentum ${\bf k}$,
$a$ is the lattice constant of the simple quadratic monolayer.
For qualitative results it is sufficient to use bulk values for the 
parameters of the paramagnetic bandstructure $\tilde{B}_i$, the crystal 
field parameter $\Delta^V_{\mbox{\footnotesize Fe/Ni}}$, and the spin
splitting parameters $J_{ex}$ and $J'_{ex}$.  
For Ni, the parameters are taken from Weling and 
Callaway~\cite{Weling82,transfni},
for Fe from Pustogowa 
{\em et al.}\/~\cite{Pusto93,transffe}. The $\tilde{B}_i$ and $\Delta^V$
are listed in the first column of table~\ref{tabkisparam}. We have used
$J_{ex}=0.1$~eV and $J'_{ex}=0.4$~eV for Ni and $J_{ex}=J'_{ex}=1.78$~eV
for Fe.
Due to the higher symmetry in fcc or bcc bulk crystals, only one
crystal field parameter $\Delta^V_{\mbox{\footnotesize Fe}}$
($\Delta^V_{\mbox{\footnotesize Ni}}$) appears in the corresponding Fe
(Ni) bulk  Hamiltonian.  For the monolayer, one would have to consider
three different $\Delta$ because of the reduced symmetry, but these
parameters are not known.  Hence,  only
$\Delta^V_{\mbox{\footnotesize Fe/Ni}}$ has been considered in Eq.~(\ref{htbfe}).
The influence of further crystal field effects on $E_{anis}$  in the
monolayer, which was stressed by Bruno~\cite{Bruno89}, 
is investigated in section III.C.

For a quantitative comparison with experiment, however,
4$s$-states have to be included 
(within the so-called ``combined interpolation
scheme''~\cite{Hodges66}) due to the strong overlap and hybridization
between 3$d$- and 4$s$-bands in 3$d$ transition metals.
According to the 
pseudopotential method by Harrison~\cite{Harrison66}, the
$4s$-electrons are described by a set of plane waves 
\[
\psi_{\K_j\k}(\r)=\frac{1}{\sqrt{Nv}}e^{i(\k-\K_j)\cdot\r}, 
\]
where the $\bf{K}_j$ are a set of reciprocal lattice vectors.
They have to be chosen such that at least the lowest eigenstates in the 
considered part of the two dimensional 
Brillouin zone (irreducible part, see below) 
are described. For simple quadratic monolayers, this yields
${\bf K}_1=(0,0)$, 
${\bf K}_2=\frac{2\pi}{a}(1,0)$, 
${\bf K}_3=\frac{2\pi}{a}(0,1)$, ${\bf K}_4=\frac{2\pi}{a}(1,1)$,
${\bf K}_5=\frac{2\pi}{a}(-1,0)$ and  ${\bf K}_6=\frac{2\pi}{a}(1,-1)$.
To maintain the symmetry of the problem (and thus the correct occurrence of 
band degeneracies that turn out to be very important for $E_{anis}$), 
{\em symmetry factors}\/ $F_i$~\cite{Hodges66} have to be introduced into the 
Hamilton matrix. This leads to
\begin{eqnarray*}
 H^s_{ij}& = &
     \left<\psi_{\K_i\k}\right|H\left|\psi_{\K_j\k}\right> \\
  & = & \left\{ \begin{array}{cl} 
  V_{00} + \alpha\left(\k-\K_i\right)^2 & \quad\mbox{for } i=j, \\
  V_{{\bf K_j}- {\bf K_i}}F_iF_j & \quad\mbox{else}
  \end{array}\right.
\end{eqnarray*}
$V_{00}$, $V_{10}$, $V_{11}$ $V_{12}$ and $V_{02}$ are the Fourier components of the 
pseudopotential, $\alpha$ is the dispersion of the 4$s$-band. The symmetry 
factors are:
\begin{eqnarray*}
 F_1 & = & 1\\
 F_2 & = & \sin 2\xi\\
 F_3 & = & \left\{\begin{array}{cl}
         \sin 2\eta & \quad\mbox{for  }  \eta \ge 0, \\
         0 & \quad \mbox{else} \\
  \end{array}\right.\\
 F_4 &=& F_2F_3\\
 F_5 & = & \left\{\begin{array}{cl}
         \sin 2\eta & \quad\mbox{for  }  \eta \le 0, \\
         0 & \quad\mbox{else} \\
  \end{array}\right.\\
 F_6 & = & F_2F_5
\end{eqnarray*} 
The $s$-$d$ hybridization $H^{sd}$ between states of parallel spins is calculated 
according to Hodges {\em et al.}\/~\cite{Hodges66} 
with the parameters $B_1$ and $B_2$.
To obtain accurate parameters, we perform a fit
to the full-potential linear muffin-tin orbitals (LMTO) calculation 
for a free-standing Fe monolayer by 
Pustogowa {\em et al.}\/~\cite{Pusto94} and 
to the linear augmented plane wave (LAPW) calculation for a Ni monolayer by Jepsen et al.~\cite{Jepsen82}. 
The resulting parameters are listed in table~\ref{tabkisparam}.
In order to to reduce the number of free parameters in the fit, 
the $d$-band parameters $\tilde{B_i}$ and $\Delta^V$ are
still taken from the corresponding bulk crystals (see above). To obtain correct
$d$-bandwidths, however, the $\tilde{B}_i$ are scaled with the fitted
parameters $S^{\uparrow}$ and $S^{\downarrow}$ for the spin-up and
spin-down bands, respectively. 
Finally, the $s$- and $d$-bandwidths and $s$-$d$-hybridization parameters 
are scaled with $t$ according to Harrison~\cite{Harrison80} to take into account the Cu 
surface lattice constant $a$:

\begin{equation}
\left(\frac{a}{a_0}\right)^{q} = \left(\frac{t}{t_0}\right)\\
\label{scaling}
\end{equation}
with $a_0$ the surface lattice constant of Fe or Ni, $t_0$ the
corresponding hopping parameters, and q being -5 for the
$dd$ parameters, -2 for the $ss$ parameters and -7/2 for
the $sd$ parameters.  The in-plane lattice constant is taken to be that of the Cu-substrate for all considered
systems ($a$ = 2.56~\AA).  This is correct for Ni, which is known to have a large
pseudomorphic growth range~\cite{Schulz94}.  For Fe however, both an
in-plane nearest-neighbor distance similar to that of Cu and a
smaller one~\cite{FB52,Mueller} have been reported.  
%
%
\subsection{Spin-orbit coupling}
Spin-orbit coupling (SOC) between the $d$-states, 
leading to magnetocrystalline anisotropy, is introduced in the usual
form as $H_{so}=\lambda_{so}{\bf l}\cdot{\bf s}$. It can be expressed by the 
components of the orbital momentum operator ${\bf l}$ in the rotated frame
$(x_M,y_M,z_M)$~\cite{Takayama76}. Here, $z_M$ is the spin quantization axis, 
which is parallel to the direction of magnetization
$(\theta,\phi)$~\cite{angles}. 
\begin{equation}
H_{so}=:\left(\begin{array}{cc} H_{so}^{\uparrow\uparrow} &
H_{so}^{\uparrow\downarrow} \\ H_{so}^{\downarrow\uparrow} &
H_{so}^{\downarrow\downarrow} \end{array}\right) = 
     \frac{\lambda_{so}}{2}
     \left(\begin{array}{cc}l_{z_M}&l_{x_M}-il_{y_M}\\
           l_{x_M}+il_{y_M}&-l_{z_M}\end{array}\right)
\label{Hso}
\end{equation}
Expressed in the basis of Eq.~(\ref{dbasis}), $H_{so}$ is a matrix function
of the magnetization direction $(\theta,\phi)$.
The SOC constant $\lambda_{so}$ is taken from the corresponding atom: 
$\lambda_{so}=70$~meV for Ni and 50~meV for Fe~\cite{Argyres55}. 

Unlike in usual tight-binding 
calculations~\cite{Bennett71,Takayama76,Bruno89,Cinal94},
SOC is included non-perturbatively~\cite{perturb} in our treatment.
Thus, we obtain important new
information on how $E_{anis}$ scales with the SOC constant $\lambda_{so}$, 
which contributes to our analysis of the origin of $E_{anis}$ 
in terms of bandstructure properties (see below). 
%
%
\subsection{Multilayers}
We build up the Hamiltonian of a system of $l$ layers by coupling $l$
monolayer Hamiltonians $H_{mono}^{i=1...l}$ together.
The coupling of the layers is described within the
tight-binding nearest-neighbor formalism used for the monolayer.
Because of the missing periodicity in {\em z}-direction, we obtain terms
that depend only on $\xi$ and $\eta$.  For the sake of simplicity, we
take only $\sigma$-bonds into account and obtain the following terms
for the coupling of the orbital j of the  monolayer i with the orbital
k of the monolayer i+1, $H^{i,i+1}_{j,k}$:
\begin{eqnarray*}
H^{i,i+1}_{22} &=& -2\tilde{B}_1\cos2\eta\nonumber\\ 
H^{i,i+1}_{33} &=& -2\tilde{B}_1\cos2\xi\nonumber
\end{eqnarray*}  
with i=1...$l$.
The ($18l \times 18l$) coupling matrix thus has only elements in the
($l-1$) (18 $\times$ 18)-blocks just above and below the diagonal.
The parameter $\tilde{B}_1$ is the same as used for the monolayers,
but it yet has to be scaled to the interlayer distance of the
tetragonally-distorted system, according to Eq. (~\ref{scaling}).  We consider
equidistant layers.
For Ni, we take into account the reported
compression of 3.2\% to scale the interlayer
hoppings~\cite{Schulz94}. For Fe, we assume an expansion of about 5\% as reported by
M\"uller $et$ $al$.~\cite{M74}.
%
%
%
\subsection{Anisotropy energy}
The magnetic anisotropy energy per atom is defined as
%
%
\begin{equation}
E_{anis}(n):=E_{tot}(\theta =0;n)-E_{tot}(\theta =\pi /2,\phi _0;n),
\label{Eanis}
\end{equation}
where $E_{tot}(\theta ,\phi ;n)$ is the ground-state energy 
per atom with a total of $n$ $3d$- and $4s$-electrons
per atom, and the magnetization direction is
denoted by $(\theta,\phi)$~\cite{angles}. 
The in-plane angle $\phi_0$ is chosen such that the resulting $|E_{anis}|$ 
is the largest possible. 
At first, the anisotropic dipole-dipole interaction 
is neglected, since it does hardly depend on the electronic structure. 
Nevertheless, it may be of the same order
of magnitude as the magnetocrystalline anisotropy resulting from SOC and 
will thus be included later to obtain quantitative results.
The total energy per atom $E_{tot}$ (with the \k-space resolved energy
$E_{\bf k}$) is given by
%
%
\begin{eqnarray}
 E_{tot}&&(\theta,\phi;n)=\frac{1}{N}
\sum_{\k} E_{\bf k}(\theta,\phi;n) \nonumber\\
 &&= \frac{1}{N}
\sum_{m,{\bf k}} E_{m{\bf k}}(\theta,\phi)
 f_0\left(E_{m{\bf k}}(\theta,\phi)-E_F(\theta,\phi;n)\right).
\label{Eanis1}
\end{eqnarray}
with N the number of atoms.
$f_0(\Delta E)$ is the Fermi-function at zero temperature and
$E_F(\theta,\phi;n)$ is the Fermi-energy which, for a given bandfilling
$n$, is determined self-consistently by
\[
n=\frac{1}{N}
\sum_{m,{\bf k}} 
 f_0\left(E_{m{\bf k}}(\theta,\phi)-E_F(\theta,\phi;n)\right).
\]
$E_{m{\bf k}}(\theta,\phi)$ is the $m$-th eigenvalue with 
crystal momentum ${\bf k}$ and magnetization along $(\theta,\phi)$
of the Hamiltonian
\[
H_{mono}=H^d+H_{so}
\]
for the monolayer in the tight-binding scheme and
\[
H_{mono}=H^{s}+H^{d}+H^{sd}+H_{so}
\]
for the monolayer in the combined interpolation scheme.
For multilayer systems, we have the following Hamiltonian:
\[
H=H_{mono}^{1} \oplus...\oplus H_{mono}^{n}+H_{coupling}
\]

In Eq. (~\ref{Eanis1}), we use the so-called {\em force theorem}, the validity of
which has been assumed in all calculations of the magnetocrystalline
anisotropy so far.

The complete Brillouin zone (BZ) 
summation over {\bf k} is performed as a weighted 
summation over the irreducible part of the BZ (for an arbitrary
direction of magnetization). For the $d$-electrons with SOC, that means a
summation over 1/4 of the BZ.  About 2000 points of the 1/4 BZ are
then sufficient to achieve convergence.  Note that we do not have to  
exclude any parts of the BZ to obtain convergence, 
unlike Wang {\em et al.}\/~\cite{wan93a}.  Adding {\em s}-electrons and
{\em s-d} hybridization implies a coupling of non-SOC coupled states with the
SOC-coupled $d$-states and results in a reduced symmetry.  It is then
necessary to perform the summation over 1/2 of the BZ.  We then need
\mbox{150 000} points to obtain the correct fourfold symmetry of the in-plane anisotropy energy
as a function of the magnetization direction in the plane
(cos4$\phi$).  Fortunately, the out-of-plane anisotropy energy
$E_{anis}$ as defined by Eq.(~\ref{Eanis}), which is larger by two
orders of magnitude in our
calculation ($E_{anis}^{in-plane} \simeq 1.2\,\mu eV$
for Fe)~\cite{Anne} already converges
for about 7000 points, so that calculations for systems of up to 14
layers are feasible.

\section{Results and Discussion}
\subsection{Monolayers within the tight-binding scheme}
In Figs.~\ref{mofe1.dat} and \ref{moni1.dat}, results for $E_{anis}$ as 
a function of the 3$d$-bandfilling $n_d$ are presented (solid lines)
for the parameters of Fe and Ni monolayers, respectively.
We use the lattice constant of 2.56~{\AA} to simulate epitaxial growth on
Cu(001). 
These figures demonstrate the correspondence between electronic structure 
and magnetic anisotropy and show that our method will yield convergent
results for the whole transition metal series and for large (Fe) and 
small (Ni) exchange coupling. They will be analyzed in the following.
Yet, the numerical value of $E_{anis}$ for Fe and Ni monolayers cannot be 
extracted from these figures until the 4$s$-electrons are included 
(see III.C), since the exact 3$d$-bandfilling of the monolayers 
is not known.

Splitting the spin-orbit coupling matrix $H_{so}$ into two parts, 
one of them ($H_{so}^{par}$) 
containing only coupling between states of parallel spin, 
the other one ($H_{so}^{antipar}$) between states of opposite spin, 
and recalculating $E_{anis}$ as a function
of $n_d$ with either of the two matrices instead of $H_{so}$ itself,
we obtain the curves $E_{anis}^{par}(n_d)$ and $E_{anis}^{antipar}(n_d)$, 
respectively (Figs.~\ref{mofe1.dat} and \ref{moni1.dat}, dashed and dotted
lines respectively).
Note that to a good approximation 
$E_{anis}^{par}(n_d)+E_{anis}^{antipar}(n_d)\approx E_{anis}(n_d)$
is valid. For Fe parameters, $E_{anis}^{antipar}(n_d)$ is very small due to
the large exchange splitting $J_{ex}$ 
that completely separates the spin subbands.
Thus, $E_{anis}^{antipar}(n_d)$ is ineffective and may therefore
be neglected for  
further analysis. The curve $E_{anis}^{par}(n_d)\approx E_{anis}(n_d)$
consists of two parts of equal shape, viz.~for $n_d\in[0;5]$ (spin-up
band) and $n_d\in[5;10]$ (spin-down band). 
In the case of Ni, $E_{anis}^{par}(n_d)$ and 
$E_{anis}^{antipar}(n_d)$ are of the same order of magnitude, since there
is a considerable overlap 
between the spin-up and spin-down subbands.

The curves $E_{anis}(n_d)$ show a number of pronounced peaks 
(A, B, C, E, F in Figs.~\ref{mofe1.dat} and \ref{moni1.dat}), the 
origin of which has to be clarified. 
Two possible contributions to $E_{anis}$ are discussed in the 
literature~\cite{Pick,wan93a,kel93}: 
(i) The SOC-induced shifting 
of occupied, nondegenerate bands leads to contributions to $E_{anis}$ in
second-order perturbation theory with respect to the SOC constant 
$\lambda_{so}$: $E_{anis}\propto\lambda_{so}^2$. 
The first order vanishes due to time reversal symmetry~\cite{Bruno89}. 
(ii) The contribution of the 
lifting of degenerate bands, which are shifted linearly with
$\lambda_{so}$, depends on the fraction
of states in ${\bf k}$-space influenced by the degeneracy.
Whether this fraction is of the order of $\lambda_{so}^2$, which would
yield~\cite{wan93a} $E_{anis}\propto\lambda_{so}^3$, or this fraction is
of lower order and thus would yield important contributions to 
$E_{anis}$,~\cite{Pick,kel93} has been a controversial question.
Anyway, the scaling of $E_{anis}$ with $\lambda_{so}$ can present important
information about the dominant contributions to $E_{anis}$.
Thus, it is very useful not to restrict calculations 
to second order perturbation theory as has been frequently done~\cite{Bruno89,Cinal94}. 
Remarkably, we find $E_{anis}(n_d)\propto\lambda_{so}^2$ for most of
the $n_d$ values 
in agreement with Wang {\em et al.}\/~\cite{wan93a}. 
Unlike stated by those authors, however, this does not rule out
contributions to $E_{anis}$  
of the lifting of degeneracies (ii). 
In section III.B and Fig.~\ref{entartung}, 
we show explicitly that such contributions play a very 
important role for $E_{anis}$ in the monolayers considered.  This is
true as well for the multilayers
(s. Figs.~\ref{band}, \ref{fe3d7_8}, and \ref{fe3dn3} and the
discussion in sections III.E and III.F).

The dependence of $E_{anis}$ on scaling of all 
$d$-electron hopping 
parameters with a common parameter $t$ was checked.  We found that
the overall shape of the curves $E_{anis}(n)$
will not change if $t$ is varied. $|E_{anis}|$
increases for decreasing $t$ (decreasing bandwidth). 
This leads to the general trend: $|E_{anis}|$ increases with
increasing lattice constant $a$ of the monolayer, since $t$ is
proportional to $a^{-5}$ (see section III.C)~\cite{Harrison80}. 
%
%
%
\subsection{The electronic origin of $E_{anis}$}
In this chapter we discuss in detail how the magnetocrystalline
anisotropy energy can be related to the electronic bandstructure.
A 3$d$-band degeneracy can make large
contributions to $E_{anis}$, if
(i) it is lifted by SOC for one direction of magnetization ($z_M^{\Xi}$)
and remains for another ($z_M^{X}$); 
(ii) it is located near the Fermi level $E_F$;
(iii) it runs along a line in {\bf k}-space and 
(iv) the degenerate bands have no or very little dispersion along this line.
Before showing that such degeneracies indeed occur in the
bandstructures, we estimate their contribution within a linearized
'bandstructure' (see Fig.~\ref{entartung}).
If $E_F$ is situated below or above the two subbands,
no contribution to $E_{anis}$ results: $\Delta
E_{anis}=0$. 
The maximal contribution occurs when the degeneracy lies exactly at the Fermi level $E_F$
and amounts to:
\begin{equation}
\Delta E_{anis}=\frac{\lambda_{so}}{2}F=
 \lambda_{so}^2\left(\frac{\partial E}{\partial k_1}\frac{\pi}{a}\right)^{-1}
\label{SPC}
\end{equation}
since the fraction $F$ of involved states in the irreducible quarter
of the BZ is $F=\frac{\Delta k_1}{\pi/a}\frac{\pi/a}{\pi/a}=
2\lambda_{so}\left(\frac{\partial E}{\partial
    k_1}\frac{\pi}{a}\right)^{-1}$.
The preferred direction of magnetization is $z_M^{\Xi}$.

Thus, $\Delta E_{anis}$ is proportional to $\lambda_{so}^2$ for a
degeneracy that 
occurs along a {\em line} with the involved bands being non-dispersive along
that line. This agrees with the scaling of $E_{anis}$ observed above.
In their estimate of the contribution of degeneracies,
Wang {\em et al.}\/~\cite{wan93a} implicitly assume that the
degenerate bands are 
dispersive in either dimension of {\bf k}-space. This would lead to
$F\propto\lambda_{so}^2$ and $E_{anis}\propto\lambda_{so}^3$ and 
justify the exclusion of degeneracies from their calculation in
order to improve convergence. In the light of our results, however, this
assumption is incorrect and it neglects very important contributions
to $E_{anis}$. 

In Fig.~\ref{figbandstr}, some degeneracies are shown in the
bandstructure of the Fe monolayer. For example, the degeneracy $A$
that occurs for ${\bf M}\parallel\hat{z}$ and is lifted for ${\bf
  M}\parallel\hat{x}$ is located at the Fermi level for $n_d=7,6$
(dotted lines in Fig.~\ref{figbandstr}) and
leads to the peak $A$
in Fig.~\ref{mofe1.dat}. It runs along a line in ${\bf k}$-space,
which is shown Fig.~\ref{line}.
According to Eq.~(\ref{SPC}),
with $\frac{\partial E}{\partial k_y} = 0.6$~eV$/\frac{\pi}{a}$ (taken
from the bandstructure), this contribution should be 
$\Delta E_{anis}\approx 4\:\mbox{meV}$, which agrees in the order of 
magnitude with the calculated value $E_{anis}(n_d=7.6)=6$~eV.

Several tests have been made to support that hypothesis. Excluding the
states influenced by the degeneracy $A$ (4.3\% of the total of 3$d$-states)
from the calculation of $E_{anis}$, the height of peak $A$ is reduced
to 40\%. The {\bf k}-space resolved analysis of $E_{anis}(n_d=7.6)$
(Fig.~\ref{kaufgeloest}) also shows clearly that $E_{anis}$ results 
from the states near the degeneracy.
 
Analogous degeneracies are found in the Ni bandstructure contributing
to the peaks $E$ and $F$ (Fig.~\ref{moni1.dat}). Note that the lifting
of degeneracies can favor in-plane as well as perpendicular
magnetization. This is in contradiction to the results of Daalderop
{\em et al.}\/ for a Co(111) monolayer~\cite{kel93},
who state that degeneracies should always favor perpendicular magnetization.

Since the 3$d$-band degeneracies are so important for $E_{anis}$, we
analyze in the following the occurrence and lifting of degeneracies in
the bandstructure. It can be shown that, 
in terms of the basis of Eq.~(\ref{dbasis}),
the Hamilton matrix $H^d$ (Eq.~(\ref{htbfe})) has the 
simplest block diagonal form with only four
off-diagonal elements (ODEs) $H^{d}_{45}=H^{d}_{54}$
and, equivalently, $H^{d}_{9,10}=H^{d}_{10,9}$.
To find out which additional
ODEs are introduced by SOC for a given direction of the magnetization
${\bf M}$, we analyze the form of $H_{so}$ in Eq.~(\ref{Hso}). 
States with 
parallel spins are coupled if they contain equal orbital momenta with
respect to the spin quantization axis $z_M$, whereas states with
opposite spins must show a difference of one in the orbital momenta to yield
nonvanishing ODEs. 
The real space components of the atomic states $\phi_i$, $i=1,...,5$, 
are composed of eigenstates of $l_z$ with the eigenvalues
\mbox{(-2,2)}, \mbox{(-1,1)}, 
(-1,1), (-2,2) and 0, respectively. In terms of eigenstates of $l_x$
one has the eigenvalues \mbox{(-1,1)}, \mbox{(-2,2)}, \mbox{(-1,1)},
(-2,0,2) and (-2,0,2), 
respectively. 
This yields a coupling for 
${\bf M}\parallel\hat{z}$ within the groups of 
states $\psi_i$ with $i=1,4,5,7,8$ and with $i=2,3,6,9,10$, and,
in the case of ${\bf M}\parallel\hat{x}$, 
within the groups of
states $\psi_i$ with $i=2,4,5,6,8$ and $i=1,3,7,9,10$, respectively.
In both cases, the Hamiltonian can be split into two $5{\times}5$ blocks,
and subbands belonging to different blocks will intersect. Between states
of the same block, the degeneracies will usually be removed.
Especially the subbands $\psi_1$ and $\psi_2$ 
(and, correspondingly, $\psi_6$ and $\psi_7$)
change their roles if the magnetization is changed from $\hat{z}$
to $\hat{x}$ and vice versa, because the orbitals $xy$ and $yz$ have
different orbital momenta with respect to the $x$- and $z$-axes. 
These subbands will thus be involved in the lifting of
degeneracies by altering magnetization and possibly, as shown above, yield
important contributions to $E_{anis}$. 
In the case of Fe parameters, 
the situation is even simpler since coupling between states
of opposite spin ($\psi_i$ and $\psi_j$ with $i\le5<j$) 
can be neglected. 

As an example, peak $A$ in
the curve $E_{anis}(n)$ of Fe at $n=7.6$ (Fig.~\ref{mofe1.dat})
results from the degeneracy $A$ (Fig.~\ref{figbandstr}) of
the subbands corresponding to the states $\psi_7$ and $(\psi_9,\psi_{10})$. 
Thus, it occurs for ${\bf M}\parallel\hat{z}$, and is lifted
for ${\bf M}\parallel\hat{x}$, since in the second case the subbands 
belong to the same block of the Hamiltonian, whereas in the first they do not.

As a conclusion, it has been shown that 3$d$-band degeneracies along lines
of constant energy result in important contributions to $E_{anis}$ if they
occur near the Fermi level. They can favor in-plane and perpendicular
magnetization and need not occur near high symmetry points of the
BZ. Thus, for (001) layers, it is not sufficient to consider only
bands at high symmetry points as was done 
by Daalderop {\em et al.}\/~\cite{kel93} for a Co(111) monolayer.
Furthermore, for such contributions from degeneracies,
$E_{anis}\propto\lambda_{so}^2$ and, approximately,
$E_{anis}\propto1/\frac{\partial E}{\partial k_1}$ is valid 
(the band
dispersion $\frac{\partial E}{\partial k_1}$ is approximately 
proportional to the scaling $t$ of the hopping parameters)
which agrees with the observations
reported above. Note that the analysis is very simple due to the
analytic form and low dimension of the 3$d$ tight-binding matrix,
which is an advantage of the semiempirical scheme.
It remains valid if the extension to $s$-states is performed (see below).
%
%
\subsection{The results of the combined interpolation scheme}
Results for $E_{anis}(n)$ obtained from 
the combined interpolation scheme (including $s$- and $d$-bands as
well as $s$-$d$ hybridization) for the monolayer are
presented in Fig.~\ref{febeslayer} for Fe parameters and Fig.~\ref{nibeslayer}
for Ni parameters with the
lattice constant of the Cu(001) surface in both cases (solid curves;  the discussion
of the curves for two and three layers is postponed to section III.E
and III.F). These results for the monolayer are similar to the curves for $d$-bands
only (Figs~\ref{mofe1.dat} and \ref{moni1.dat}). $n$ is the total filling of
the $s$- and $d$-band ($n=8$ for Fe and $n=10$ for Ni).
We find for a Fe monolayer
$E_{anis}(\mbox{Fe/Cu})=-0.41$~meV per atom, for Ni
$E_{anis}(\mbox{Ni/Cu})=0.10$~meV per atom. The 
dipole-dipole interaction is included under the assumption of
a point dipole located at each site, carrying the magnetic
moment of the unit cell. The (spin) magnetic moment per
atom is calculated from the bandstructure
($m(\mbox{Fe/Cu})=3.3\mu_B$
and $m(\mbox{Ni/Cu})=0.91\mu_B$). The dipole anisotropy
(equivalent to the shape anisotropy in the monolayer) always prefers
in-plane magnetization.
Alltogether, we obtain for the total magnetic anisotropy energy per
atom of a Fe and
Ni monolayer with the lattice constant of Cu(001)
\begin{eqnarray*}
E_{anis}^{tot}(\mbox{Fe/Cu}) &=&-0.17\:\mbox{meV} \quad\mbox{and}\\
      E_{anis}^{tot}(\mbox{Ni/Cu})&=&\mbox{ 0.12}\:\mbox{ meV} 
\end{eqnarray*}
with the easy axis perpendicular to the monolayer for Fe and in-plane for Ni.
Note that corresponding {\em ab-initio}\/ results for a free-standing
Fe-monolayer  
yielded $-0.42$~meV~\cite{Gay86,wan93a}, 
but previous tight-binding calculations gave the too large value of 
$-5.5$~meV~\cite{Cinal94}. 

In the case of Fe, the perpendicular easy axis of ultrathin Fe films
on Cu(001) is reproduced correctly. Direct comparison with a Fe
monolayer on Cu(001) is difficult due to film growth
problems~\cite{expts}.
It is common use to separate the anisotropy energy  of thin films in a volume and a
surface term~\cite{FB53,Schulz94}:
\begin{equation}
E_{anis}(d)={\bf K}_v+\frac{2{\bf K}_s}{d},
\label{kv}
\end{equation}
The first term, ${\bf K}_v$, describes the thickness independent
contributions to the anisotropy energy, and the second, ${\bf K}_s$, the thickness dependent
contributions and the surface effects. 
Fowler and Barth measure the following anisotropy constants~\cite{FB53}:
\mbox{${\bf K}_v$ = 0.132 meV/atom} and \mbox{${\bf K}_s$ = 0.11
  meV/atom} for the distorted fcc-films at 100 K.  The value
\mbox{${\bf K}_v$+2${\bf K}_s$ = 0.352 meV/atom} is comparable to our result.
This result has been calculated with the measured anisotropy field
using the bulk saturation magnetization of bcc-Fe.
For Ni, our result also agrees very well with experiments~\cite{Schulz94} which yields
$E_{anis}(\mbox{Ni/Cu})=0.125$~meV at 300 K.
The anisotropy constants ${\bf K}_s$ and ${\bf K}_v$ are temperature
dependent.  Measurements of the anisotropy
constants as a function of the reduced temperature have been
made~\cite{Farle}, but the correct extrapolation to T=0 K is not known
yet.  While in experiment, the values of $K_{v}$ and $K_{s}$ have to
be compared at the same reduced temperature because of the thickness
dependence of $T_c$, the theoretical values are for 0 K and thus
independent of the difference of absolute and reduced temperature.

Note that in
Fig.~\ref{nibeslayer}, the curve $E_{anis}(n)$ for the Ni monolayer
(solid curve)
has zeros near $n=10$. Hence, the numerical result for Ni is not
very stable and the excellent agreement with experiment should not be
overemphasized. Nevertheless, for Fe and Ni, the sign and the order of
magnitude of $E_{anis}$ turn out to be remarkably
stable upon parameter variations:
Sign changes do not occur upon variation of the
pseudopotential and $s$-$d$ hybridization parameters by as much as 40\%.
Moreover, we find in agreement with Wang {\em et al.}\/~\cite{wan93a} a
perpendicular easy axis also for Fe monolayers taking (001) surface
lattice constants imposed by substrates such as 
Pd, Ag, and V (2.77~{\AA}, 2.89~{\AA}, and 3.03~{\AA}), respectively.
This stability again demonstrates the validity of our results for
$E_{anis}$. The good agreement of the results both with {\em ab-initio}\/
theories and experiments is due to the fact that the parameters were
obtained by a fit to {\em ab-initio}\/ calculations for Fe and Ni {\em
monolayers}\/ rather than taking {\em bulk}\/ parameters.

To investigate crystal field effects, an additional parameter $\Delta$
is introduced~\cite{Bruno89} to take into account the different effect of the
monolayer geometry on orbitals that lie in the plane of the monolayer
($xy$ and $x^2-y^2$) 
and out-of-plane orbitals ($yz$, $zx$ and $3z^2-r^2$). 
Additional to Eq.~(\ref{htbfe}), the on-site energies of
the latter are lowered by $\Delta$ with respect to the first. 
The dependence of $E_{anis}$ on $\Delta$ is shown in Fig.~\ref{figkfa}
for Fe and Ni parameters (solid and dashed curve,
respectively). Remarkably, $\Delta=0.2$~eV changes the sign of
$E_{anis}$ for both systems considered. Thus, it is important to
determine $\Delta$ from the {\em ab-initio}\/ bandstructures. In the
case of Fe, the fit of the 3$d$-bands near the
$\overline{\Gamma}$-point of the BZ can be significantly improved by
chosing $\Delta=0.08$~eV. The resulting $E_{anis}(\mbox{Fe/Cu})$ amounts to
$-0.30$~meV, still with perpendicular easy axis even if the
dipole-dipole interaction is added.
For Ni, the introduction of $\Delta$ does {\em not}\/ improve the fit.  
Those results for $\Delta$ differ substantially from $\Delta=-0.5$~eV given by
Bruno~\cite{Bruno89} which has been determined by a fit to the Ni(111)
monolayer but employed for both 
Fe and Ni(001) monolayers also. Pick and Dreyss\'e~\cite{Pick}
state that for (001)-monolayers a crystal field parameter is not
necessary. For Ni, this is supported by our result; even in Fe, our
value of $\Delta$ is small compared to other bandstructure parameters.
Cinal {\em et al.}\/~\cite{Cinal94} report $\Delta=-0.14$~eV for the
Ni(001) monolayer.

Finally, a detailed investigation of the bandstructures~\cite{MoosDA}
shows that the analysis given in section III.B for 3$d$-bands only is
still valid for the combined interpolation scheme. As an evidence,
consider Figs~\ref{mofe1.dat} and \ref{febeslayer} (solid curves): There
is a one-to-one correspondence between the 
peaks in $E_{anis}$ in both curves.  This correspondence can be shown to result from
similar bandstructure 
details. In particular, the role of 3$d$-band degeneracies stressed in
section III.B remains the same in the complete scheme.
\subsection{Temperature dependence}
One of the greatest challenges in the investigation of magnetic
anisotropy is the calculation of reorientation transitions with
temperature. Up to now, a complete electronic and thermodynamic theory
is lacking. Here, one-particle effects of temperature are
investigated. It turns out that they again support
the role of degeneracies for magnetic anisotropy and, moreover, are
comparable in order of magnitude with the many-particle aspects
usually considered~\cite{Jensen90}. 

The free magnetic anisotropy energy $F_{anis}$ depends on temperature
$T$ due to (i) the Fermi distribution of electronic states $f_T(\Delta
E)$, (ii) the hopping integrals, which depend on T because of the lattice
expansion of the substrate, (iii) the entropy $S(T)$ and (iv) the
effects of spin-waves, resulting in a temperature dependence of the
magnetization ${\bf M}(T)$. In this work, the first three effects are
analyzed. More precisely, the thermal expansion (ii) of the lattice
constant $a(T)$
is included by means of the empirical law
$a(T)=a(T=0)(\alpha T+1)$. $\alpha=2\cdot10^{-5}/K$ is the expansion
coefficient for the Cu substrate~\cite{Collieu73}. 
The expression for the entropy (iii) of non-interacting particles is:
\[
S=-k_B\sum_{m,\k}\left<n_{m\k}\right>\ln\left<n_{m\k}\right>+
       (1-\left<n_{m\k}\right>)\ln(1-\left<n_{m\k}\right>)
\] 
with $\left<n_{m\k}\right>=f_T\left(
E_{m\k}(\theta,\phi)-\mu(\theta,\phi;n)\right)$.
In analogy to Eq.~(\ref{Eanis}), the free magnetocrystalline anisotropy
energy $F_{anis}$ is defined as the difference in the free energy $F=E-TS$ for
two different directions of magnetization.

Fig.~\ref{figtemp} shows $F_{anis}(T)$ ($d$-band calculation for the monolayer, Fe
parameters, $n_d=6$). Including only Fermi statistics (i; dashed
curve), the characteristic energy scale for the decrease
of $|F_{anis}|$ with $T$ is about 1000~K (100~meV), which corresponds to
the energy $2\lambda_{so}$, but not to the 3$d$-bandwidth of
approximately 3~eV. This becomes immediately plausible if one notices
that the SOC-induced lifting of degeneracies occurs near the Fermi
level. Thus, one expects a measurable effect on $F_{anis}$ 
due to Fermi statistics as soon as $k_BT$
becomes larger than or comparable to $2\lambda_{so}$. In addition, we
must conclude from our results that shifting of 
subbands far below the Fermi-level is not so important, since 
then $F_{anis}$ could not be essentially lowered on such a 
small temperature scale.

The characteristic increase of $|F_{anis}|$ with increasing temperature
for $T<500$~K is a direct result of the lifting of
degeneracies. Consider again Fig.~\ref{entartung}: for ${\bf
  M}\parallel z_M^{\Xi}$ (lifted degeneracy), which is the
energetically favored case, 
Fermi statistics induces only little changes in
the occupation of the electronic states, if $k_BT<\lambda_{so}$;
for the degenerate bands (${\bf M}\parallel z_M^{X}$), however, 
states in the upper band are significantly
occupied even for $k_BT<\lambda_{so}$. 
Thus, the total energy for ${\bf M}\parallel z_M^{X}$
rises with respect to $T=0$ in this temperature range. This leads
to an increase of $|F_{anis}|$ with increasing $T$, if
$k_BT<\lambda_{so}=50$~meV ($T<500$~K).

The inclusion of lattice expansion (ii; solid curve in
Fig.~\ref{figtemp}) has only a small effect on $F_{anis}$. 
The narrowing of bands with increasing temperature due to the scaling
of the hoppings leads to an increase in $|F_{anis}|$ {\em for small} $T$,
which was already discussed for $T=0$.
{\em For larger} $T$, the influence of Fermi statistics on narrowed bands is
larger, leading to a stronger decrease of $|F_{anis}|$. 

The entropy (iii; dotted curve) has a damping effect on the curve 
$F_{anis}(T)$, but
maintains the features discussed above. This results from the fact
that, in the case of degenerate bands, the entropy is larger than for
nondegenerate bands, since states located nearer to the Fermi level
have larger entropy.

$F_{anis}(T)$ was also calculated for bilayers (Fig.~\ref{templ2}),
taking into account all three mentioned effects
and shows a decrease with increasing temperature on the same scale as for the
monolayer.  Hence, this analysis of $F_{anis}(T)$ shows the significant contribution
of temperature-induced changes of the degeneracies to the
anisotropy energy. It is remarkable that the three temperature effects
mentioned above, and particularly the electron temperature dependence of
the Fermi-function, are of equal magnitude as the temperature effects of
spin-waves on ${\bf M}(T)$. 
%
%
\subsection{Fe Multilayers}
Fig.~\ref{fe(l)} shows the calculated magnetic anisotropy energy for Fe films
of 1 to 14 layers.  Calculations for both 1/4 BZ and 1/2 BZ are
included.  The values obtained when the summation over {\bf k} is
performed over 1/4 BZ lead to periodically recurring positive values
of $E_{anis}$ (for films of 2, 6, 9 and 12 layers).  The positive
value for film of 2 layers can be traced back to the occurence of 
degeneracy A at the Fermi-level.  For the other positive values, easy
and hard axis are found to be in-plane, an effect of the wrong symmetry
resulting from the summation over 1/4 BZ, and the following
overestimation of the {\em in-plane} anisotropy ${\bf  E}_{anis}^{in-plane}$.
 A new degeneracy D is found to
be responsible for the negative values. In Fig.~\ref{band}, we show the
band structure of the Fe monolayer calculated within the combined
interpolation scheme.  The degeneracy A observed for the monolayer in the
tight-binding scheme is easy to recognize,
and a new degeneracy D is found near the M-point for
${\bf M}\parallel\hat{x}$.  
Degeneracy D is lifted for ${\bf M}\parallel\hat{z}$, thus leading to a
negative anisotropy energy.  The {\bf k}-space
resolved anisotropy energy shown in Fig.~\ref{fe3d7_8} confirms the
importance of degeneracy D, which causes the ring-shaped dip around M.
The structure seen along the line LL' (s. Fig.~\ref{line}) is the onset
of the positive peak in the anisotropy energy caused by degeneracy A
(s. Fig.~\ref{febeslayer}).  
Summation of the contributions of the {\bf k}-points in the tenth of the BZ
near M already gives half of the total anisotropy energy.  Multilayer
systems show per se more degeneracies than monolayers, and the contribution of
these to the total anisotropy energy is not as clear as for the
monolayer.  Still, for a three layer
system, we find again degeneracy D at the Fermi level, and recognize
in  the {\bf k}-space resolved anisotropy energy (Fig.~\ref{fe3dn3})
the characteristic structure it causes around M.  
We perform the {\bf k}-space summation again, this time over 1/2 BZ, thus
respecting the symmetry of the s-d hybridized system. This reduces the importance of
degeneracy A and we find positive values only at 2 and 6 ML.  Taking
Fig.~\ref{fe(l)} again and excluding the points of wrong symmetry
(easy and hard axis in-plane) and the points
where we find degeneracy A at the Fermi-level, we obtain the curve
shown in Fig.~\ref{fe(1/l)}.  ${\bf E}_{anis}$ is plotted as a function of
1/{\em l}.  We expected a linear behavior (s. Eq. (~\ref{kv})) and thus
performed a linear least-square fit to the data.  
We then obtain \mbox {${\bf K}_v$ = -0.17 meV per atom} and \mbox{${\bf
      K}_s$ = -0.28 meV per atom}, which is in very good agreement
with Fowler und Barth~\cite{FB53}. 

Including the dipole-dipole anisotropy energy as calculated by
L. Szunyogh {\em et al.}~\cite{Weinberger}, we would expect for Fe a change
of the easy-axis from perpendicular to in-plane at 4 ML
\mbox{($E^{dip}_{anis}$(4 layers)=0.59meV)}. Our result indicates that
the experimentally observed transition at 5 ML might be an intrinsic
quality of fct-films grown at low temperature.   L. Szunyogh {\em et  al.} calculated
the anisotropy energy of thin fcc-Fe films on Au (001) and also observed
oscillations.  They obtained a reorientation transition from
perpendicular to in-plane magnetization at 4 ML.

The dependence of $E_{anis}$ of Fe on the 3$d$- and 4$s$-bandfilling n is
shown in Fig.~\ref{febeslayer}.  For the monolayer at a $s$- and $d$-bandfilling n=8, we are near a
zero of the curve, and at n=8.2 we already have a positive
value of $E_{anis}$ caused by the growing influence of degeneracy A.
We would thus expect a monolayer of a $Fe_xCo_{1-x}$  alloy to have an
in-plane magnetization already at small Co concentrations.  This was
in fact measured by Dittschar {\em et al}~\cite {Kirschner} for x=0.95.  We would
predict an increase of the anisotropy energy with increasing Co
concentration.  For 3 layers, we would expect the same behavior, the
structure of the curve ${\bf E}_{anis}$(n) near n=8 being similar to that of the monolayer.
This alloying behavior found both theoretically  and experimentally supports
the relevance of degeneracies for the anisotropy energy, as claimed by
Daalderop {\em et al.} and disputed by Wang, Wu and Freeman.  In this case,
there is no doubt that the magnetic moment persists.

\subsection{Ni Multilayers}
The magnetic anisotropy of Ni calculated for systems of 1 to 14 ML is
shown in Fig.~\ref{nibeslayer}.  We include again calculations using 1/4 of the
BZ and 1/2 BZ, but this time no point has to be excluded. ${\bf
  E}_{anis}$ of the second layer is much bigger than that of the
monolayer, a fact which indicates that the influence of the
substrate maybe cannot be neglected.  The anisotropy then sinks again and
remains approximately constant at a value of about 0.14 meV (which is
still bigger as the value obtained for the monolayer).  Schulz and
Baberschke~\cite{Schulz94} report for Ni a transition from in-plane to perpendicular
magnetization at 7 ML, due to a large ${\bf K}_v$ which favors
perpendicular orientation of the magnetization.  Our theory does not
reproduce this reorientation.  

For Fe, the behavior of the films as a function of thickness could be
related to the degeneracies occurring at the Fermi-level.  The contribution
of these  degeneracies to the total anisotropy of the film would be
expected to decrease with increasing number of layers, as their weight
in the summation over all atoms (number of points in the BZ $\times$
number of layers) decreases: that is in fact what we find for Fe.   For
Ni, the contribution of the degeneracies to the anisotropy energy is
not so evident, the minority and majority spin bands mix much more than in the case of Fe because of the small
exchange coupling.  This is a possible reason for the nearly constant
anisotopy energy we obtain.  The occurrence of a degeneracy of a {\em
  l}-times degenerated band would also probably lead to a thickness
independent contribution to ${\bf E}_{anis}$.

So far, no other monolayer calculation lead to the correct in-plane
anisotropy for the Ni-monolayer.  In a calculation for the fct-bulk,
Eriksson~\cite{Erik2} finds a perpendicular easy-axis, which is correct
for fct-Ni, but wrong for
fcc-Ni.  We obtain the correct in-plane anisotropy for the monolayer,
but the wrong ${\bf K}_v$.  So, a 3D calculation for fct-Ni does not
really tackle the problem and explain the behavior of the
magnetization.

Ni is a delicate system.  Maybe many-body effects cannot be neglected
(i.e. the {\em force theorem} does not work well).  However, a
total-energy calculation made by Eriksson~\cite{Eriksson} still yields the wrong sign
for ${\bf E}_{anis}$ for fcc-Ni bulk. The dependence between
the anisotropy energy and the bandstructure seem to be very subtle
and the smallest details can influence the results. 

%
\section{Conclusions}
A calculation of the magetocrystalline anisotropy
energy $E_{anis}$ of Fe and Ni monolayers on Cu(001) is performed. 
In agreement with experiments, we find a perpendicular easy axis 
for Fe and an in-plane easy axis for Ni. 
The results are fully converged without any additional assumption to
improve convergence. SOC is included non-perturbatively.
It is an important result that large contributions to $E_{anis}$
can result from the SOC-induced lifting of degeneracies occuring along
{\em lines}\/ in ${\bf k}$-space at the Fermi-level.  
The contributions of those degeneracies scale with the square of
the SOC constant $\lambda_{so}$, as contributions from nondegenerate
bands do. 
The occurrence and lifting of degeneracies in the 3$d$-band has been
discussed in general. 
Evidence for the important contribution to ${\bf E}_{anis}$ of the degeneracies at the
Fermi-level are (i) the groove and the
ring-shaped dip in the {\bf
  k}-space resolved anisotropy for the monolayer in the tight-binding
scheme and in the combined interpolation scheme respectively, (ii) the
temperature dependence (the characteristic energy scale for the
decrease of the free magnetocrystalline anisotropy energy $|F_{anis}|$
as a function of the temperature is determined by $\lambda_{so}$),
(iii) the finite  anisotropy energy at ${\bf T}_c$ and (iv) the
alloying behavior of $Fe_xCo_{1-x}$.
We obtain for Fe a reorientation transition from perpendicular to
in-plane magnetization at 4 ML, reorientation which is independent of any
restructuration of the fct-film.
Since it can be seen from Fig.~\ref{febeslayer} and Fig.~\ref{nibeslayer} that both Fe
and Ni do not exhaust the maximal anisotropy possible, 
our calculation of $E_{anis}$ should also be important for the
technologically relevant maximization of magnetic anisotropy by
appropriate surface-alloy formation.  

\newcommand{\PRL}{Phys. Rev. Lett.~}
\newcommand{\PRB}{Phys. Rev. B~}
\newcommand{\PR}{Phys. Rev.~}

\end{multicols}
\newpage
%
%
\begin{twocolumn}
\begin{table}[tb]
\caption{Bandstructure parameters within the combined interpolation
  scheme for Fe and Ni (001)-monolayers with
  lattice constant $a$.
  The parameters $\tilde{B}_i$ and $\Delta^V$ are taken from Pustogowa
  {\em et al.}$^{28,29}$  for Fe and from Weling and Callaway$^{26,27}$
  for Ni (bulk parameters).  
  The other parameters are obtained from a fit to {\em ab-initio}\/
  calculations for freestanding (001)-monolayers 
  by Pustogowa {\em et al.}$^{31}$
  for Fe and Jepsen  {\em et al.}$^{32}$  for Ni.}
\vspace{.2cm}
\centering
\label{tabkisparam}
\begin{tabular}{cr@{}lr@{}l}
 & &Fe & &Ni \\ \hline
$\quad\tilde{B}_1$(eV)$\quad$ & 0&.0774 & 0&.152923 $\quad\quad$\\ 
$\tilde{B}_2$ (eV) & -0&.00816 & -0&.015135 \\ 
$\tilde{B}_3$ (eV) & 0&.0774 & 0&.227635 \\ 
$\tilde{B}_4$ (eV) & -0&.15324$\quad\quad$ & -0&.25 \\ 
$\tilde{B}_5$ (eV) & -0&.05652 & -0&.071149 \\ 
$\tilde{B}_6$ (eV) & 0&.08376 & 0&.119380 \\ 
$\Delta^V$ (eV)  
       & 0&.068 & 0&.059360 \\ \hline
$\quad S^{\uparrow}\quad$ & 2&.06 & 1&.33 \\
$S^{\downarrow}$ & 2&.63$\quad$ & 1&.52$\quad$ \\ 
$J_{ex}$ (eV) & 2&.18 & 0&.87 \\
$J'_{ex}$ (eV) & 2&.18 & 1&.17 \\
$E_0$ (eV)    & -0&.54 & -0&.935 \\ \hline
$\alpha$ (eV) & $\quad$20&.0  & $\quad$25&.2  \\
$V_{00}$ (eV) & -4&.20 & -4&.60 \\
$V_{10}$ (eV) &  1&.2  &  0&.4  \\
$V_{11}$ (eV) &  1&.0  &  2&.0  \\ \hline
$B_1$ (eV)    &  7&.5  &  5&.0  \\
$B_2$ (eV)    &  5&.1  & 12&.8  \\ \hline
$a$ (\AA) & 2&.76 & 2&.49 \\      
\end{tabular}
\end{table}
%
\begin{figure}
\centerline{\psfig{figure=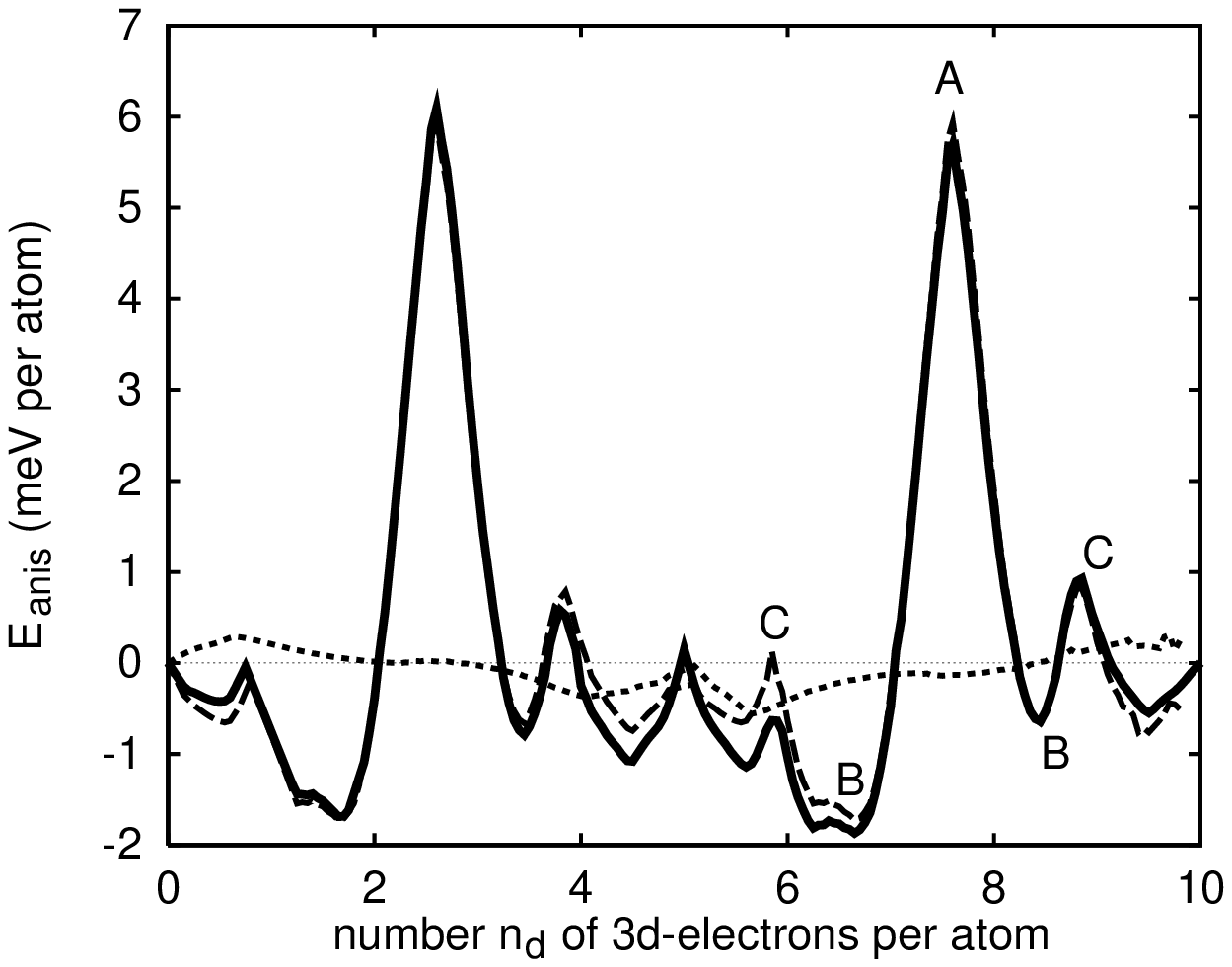,width=8.50cm}}
\vspace{.3cm}
\caption{Dependence of the magnetocrystalline anisotropy energy
$E_{anis}$ on the $3d$-bandfilling $n_d$
for a monolayer with parameters referring to Fe, 
calculated within the tight-binding scheme (solid curve).
Negative values of $E_{anis}$ yield perpendicular anisotropy.
The origin of the peaks denoted by $A$, $B$ and $C$ can be traced back to
degeneracies in the bandstructure (see text and Fig.~\ref{figbandstr}). 
The dashed and dotted curves show the contributions $E_{anis}^{par}$
and $E_{anis}^{antipar}$ to $E_{anis}$ from
the spin-orbit coupling between parallel spins and antiparallel spins,
respectively.} 
\label{mofe1.dat}
\end{figure}
\begin{figure}
\centerline{\psfig{figure=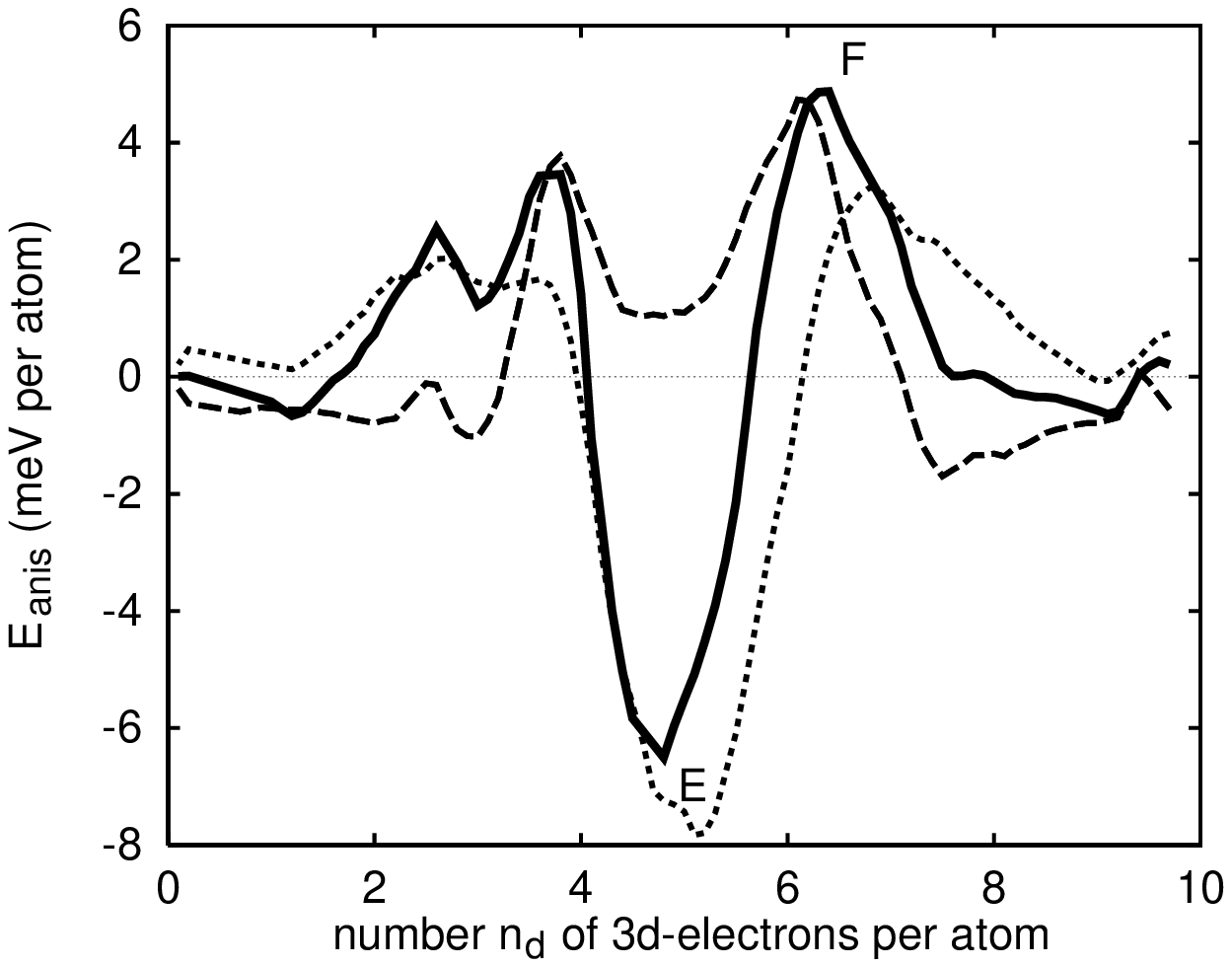,width=8.50cm}}
\vspace{.3cm}
\caption{Dependence of the magnetocrystalline anisotropy energy
$E_{anis}$ on the $3d$-bandfilling $n_d$
for a monolayer with parameters referring to Ni, 
calculated within the tight-binding scheme (solid curve).
Negative values of $E_{anis}$ yield perpendicular anisotropy.
The origin of the peaks denoted by $E$ and $F$ can be traced back to
degeneracies in the bandstructure (see text). 
The dashed and dotted curves show the contributions $E_{anis}^{par}$
and $E_{anis}^{antipar}$ to $E_{anis}$ from
the spin-orbit coupling between parallel spins and antiparallel spins,
respectively.} 
\label{moni1.dat}
\end{figure}
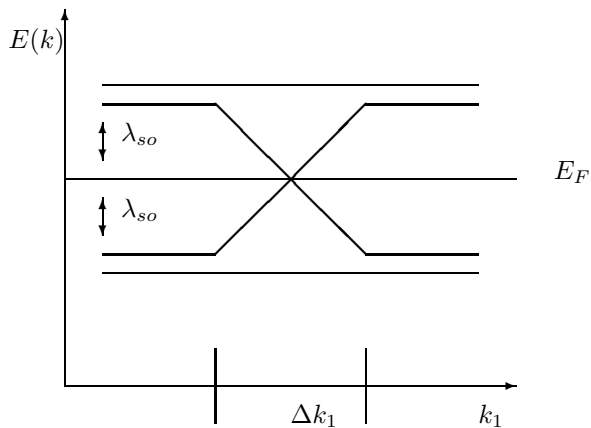
\begin{figure}
\vspace{.3cm}
\unitlength0.5cm
\begin{picture}(12,12)
\thinlines
\put(2,2) {\vector(1,0){12.0}}
\put(2,2) {\vector(0,1){10.0}}
\put(3,10) {\line(1,0) {10.0}}
\put(3,5) {\line(1,0) {10.0}}
\put (2,7.5) {\line(1,0) {12.0}}
\put (15,7.5) {\bf $E_F$}
\put(13,1) {{\bf $k_1$}}
\put(0.5,11) {{\bf $E(k)$}}
\put(6,1) {\line(0,1) {2.0}}
\put(10,1) {\line(0,1) {2.0}}
\put(8,1) {{\bf $\Delta k_1$}}
\thicklines
\put(3,9.5) {\line(1,0) {3.0}}
\put(6,9.5) {\line (1,-1) {4.0}}
\put(10,5.5) {\line(1,0) {3.0}}
\put(3,5.5) {\line(1,0) {3.0}}
\put(6,5.5) {\line(1,1) {4.0}}
\put(10,9.5) {\line(1,0) {3.0}}
\thinlines
\put(3,6.0) {\vector(0,1) {1.0}}
\put(3,6.5) {\vector(0,-1) {0.5}}
\put(3,8.0) {\vector(0,1) {1.0}}
\put(3,8.5) {\vector(0,-1) {0.5}}
\put(3.5,6.5) {$\lambda_{so}$}
\put(3.5,8.5) {$\lambda_{so}$}
\end{picture}
\caption{Occurrence (full lines) and lifting (dashed lines) of a ``line'' 
degeneracy for two different
directions of magnetization $z_M^X$ and $z_M^{\Xi}$, respectively. 
${\bf k}_1$ corresponds to one particular direction in {\bf
k}-space. Perpendicular to ${\bf k}_1$ the intersecting bands are
non-dispersive throughout the BZ.
Note, the energy gained by the
lifting of this degeneracy is given by $\Delta E_{anis}=\frac{1}{2}
\lambda_{so}\cdot F$, if $E_F$ falls in between the two subbands 
(dotted line). Here, $F$ is the fraction of the involved states in 
{\bf k}-space. Apparently, if $E_F$ lies
below or above the two subbands, $\Delta E_{anis}$ is zero.}
\label{entartung}
\end{figure}
\begin{figure}
\centerline{\psfig{figure=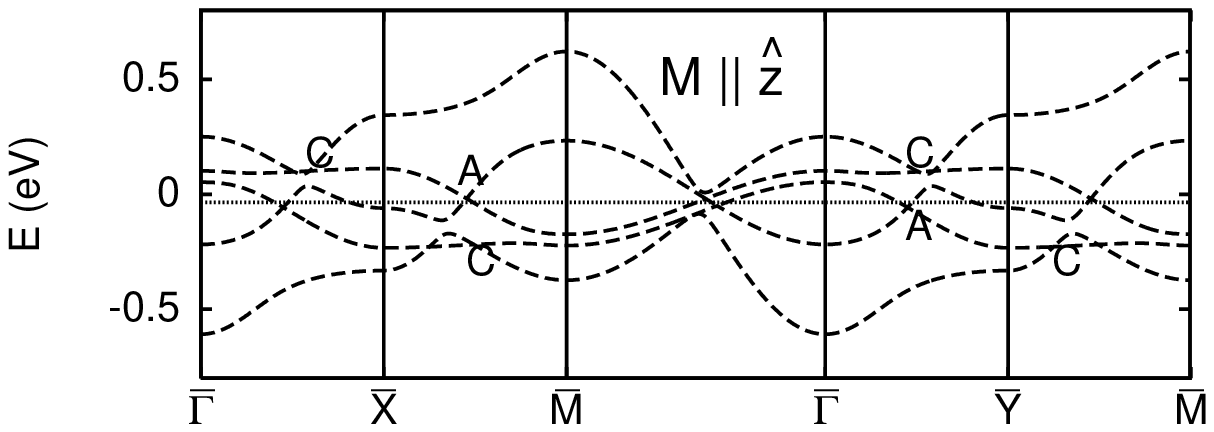,width=8.50cm}}
\vspace{.3cm}
\centerline{\psfig{figure=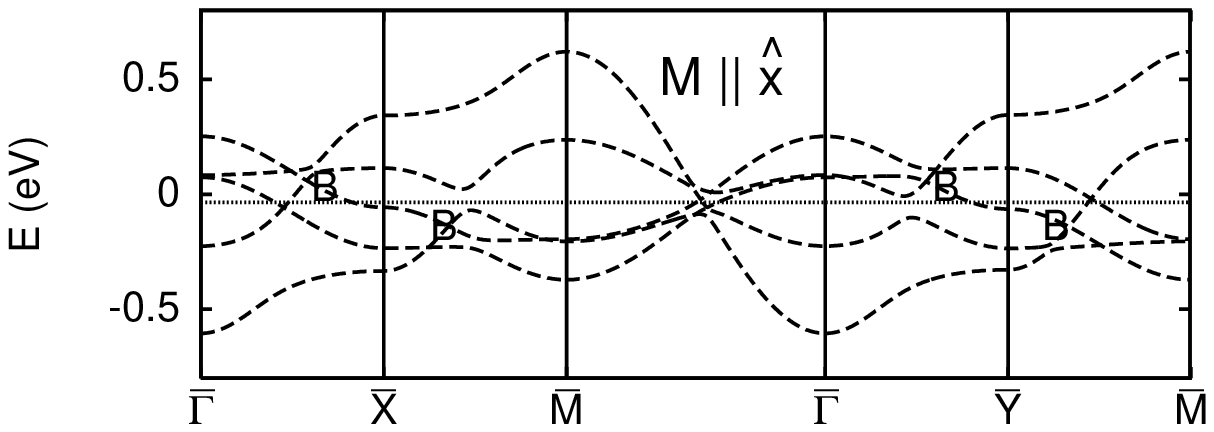,width=8.50cm}}
\vspace{.3cm}
\caption{
Bandstructure of the 3$d$ minority spin band of the Fe monolayer,
calculated within the tight-binding scheme. The magnetization ${\bf
  M}$ is
directed along the layer normal $\hat{z}$ (upper part) and in-plane
along $\hat{x}$ (lower part). The degeneracies denoted by $A$, $B$ and $C$
contribute to the peaks $A$, $B$ and $C$ in Fig.~1. The dotted lines
denote the Fermi level for $n_d=7,6$,
respectively. $\overline{\Gamma}=(0,0)$, $\overline{X}=(\pi/a,0)$,
$\overline{Y}=(0,\pi/a)$ and $\overline{M}=(\pi/a,\pi/a)$ are the high
symmetry points of the irreducible part ($0\le k_x,k_y\le\pi/a$) 
of the Brillouin zone. $a$ is the lattice constant of the monolayer.}
\label{figbandstr}
\end{figure}
\begin{figure}
\centerline{\psfig{figure=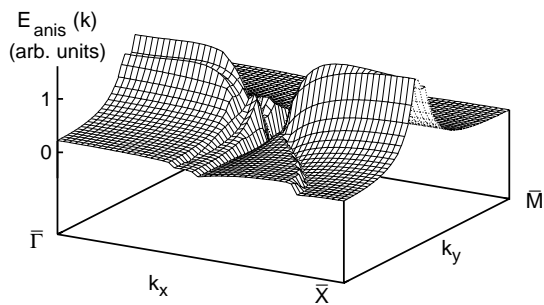,width=8.50cm}}
\vspace{.3cm}
\caption{\k-space resolved magnetocrystalline anisotropy energy
  $E_{anis}(\k,n_d)=E_{\k}(\theta=0;n_d)-E_{\k}(\theta=\pi/2,\phi=0;n_d)$
  in the irreducible part of the Brillouin zone
  ($0\le k_x,k_y\le\pi/a$; $a$ is the lattice constant of the
  monolayer) for the bandstructure of the Fe monolayer,
  $n_d=7.6$ electrons per atom in the $d$-band, calculated within the
  tight-binding scheme. Positive values of $E_{anis}(\k,n_d)$ favor
  in-plane magnetization.}   
\label{kaufgeloest}
\end{figure}
\begin{figure}
\centerline{\psfig{figure=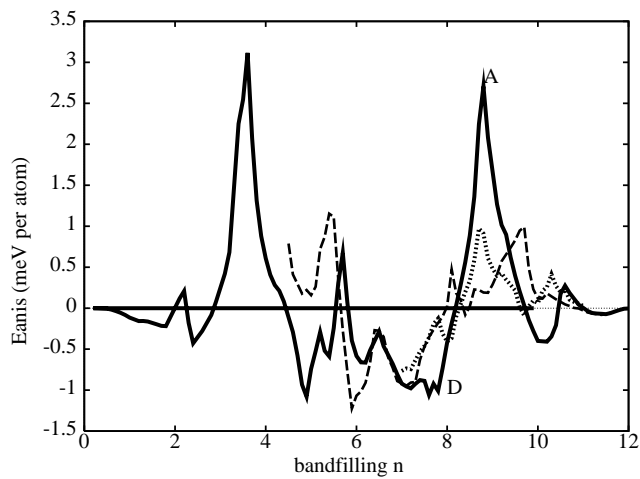,width=8.50cm}}
\vspace{.5cm}
\caption{Magnetic anisotropy energy of Fe as a function the $s$- and
$d$-bandfilling for 1 layer (solid curve), 2 layers (dashed) and 3
layers (dotted).  Peaks A and D are caused by the respective
degeneracies in the bandstructure shown in Fig.~\ref{band}.}
\label{febeslayer}
\end{figure}
\begin{figure}
\centerline{\psfig{figure=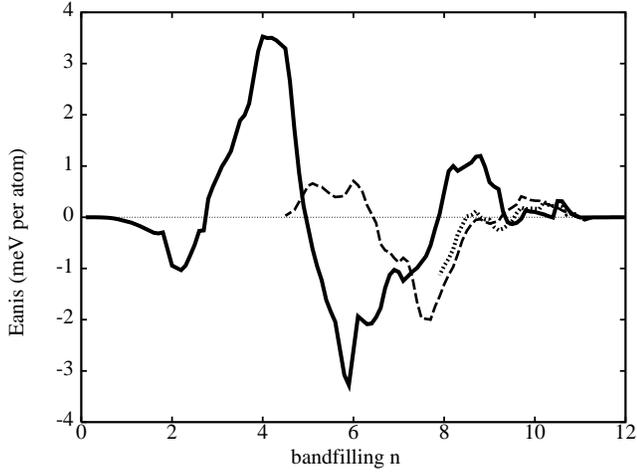,width=8.50cm}}
\vspace{.5cm}
\caption{Magnetic anisotropy energy of Ni as a function the $s$- and
$d$-bandfilling for 1 layer (solid curve), 2 layers (dashed) and 3
layers (dotted).}
\label{nibeslayer}
\end{figure}
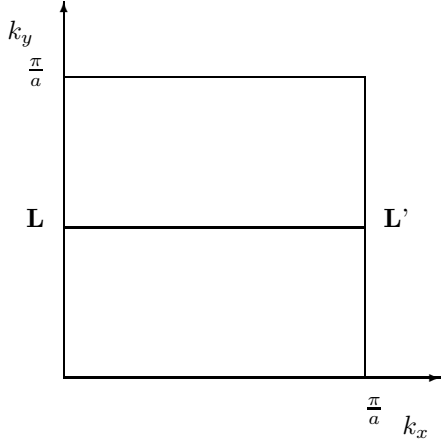
\begin{figure}
\vspace{.5cm}
\unitlength0.5cm
\begin{picture}(12,12)
\thinlines
\put(2,2) {\vector(1,0){10.0}}
\put(2,2) {\vector(0,1){10.0}}
\thicklines
\put(2,6) {\line(1,0) {8.0}}
\thinlines
\put(2,10){\line(1,0) {8.0}}
\put(10,2){\line(0,1) {8.0}}
\put(11,0.5) {{\bf $k_x$}}
\put(0.5,11) {{\bf $k_y$}}
\put(1,10) {{\bf $\frac{\pi}{a}$}}
\put(10,1) {{\bf $\frac{\pi}{a}$}}
\put(1,6) {{\bf L}}
\put(10.5,6) {{\bf L}'}
\end{picture}
\caption{Irreducible part of the two-dimensional Brillouin zone of Fe
  for the tight-binding scheme. 
  $a$ is the lattice constant of the monolayer. The
  main contribution to $E_{anis}$ at $n=8.8$ (corresponding to
  $n_d=7.6$ in the tight-binding calculation) results from the lifting
  of degeneracies along the line $LL'$. }
\label{line}
\end{figure}
\begin{figure}
\centerline{\psfig{figure=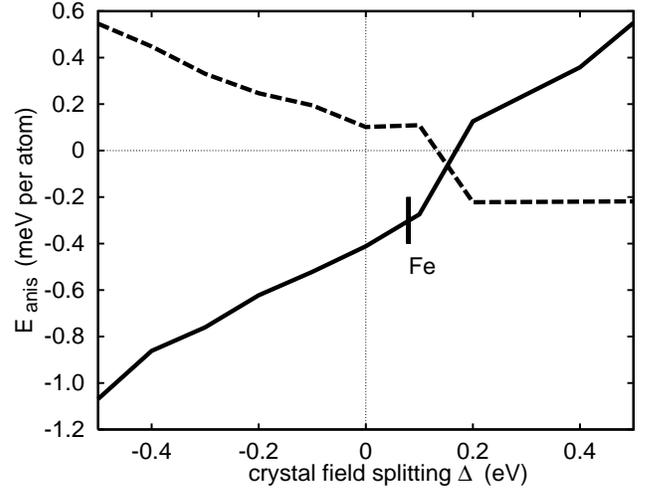,width=8.50cm}}
\vspace{.5cm}
\caption{Dependence of the magnetocrystalline anisotropy energy
$E_{anis}$ on the crystal field splitting $\Delta$ for the Fe 
monolayer on Cu(001), $n=8$, (solid curve) and the Ni monolayer on Cu(001),
$n=10$, (dashed curve). Negative values of $E_{anis}$ yield
perpendicular anisotropy. The vertical line denotes the best fit for
$\Delta$ for the Fe monlayer. In the case of Ni, the fit cannot be
improved by the introduction of $\Delta$ (see text).}
\label{figkfa}
\end{figure}
\begin{figure}
\centerline{\psfig{figure=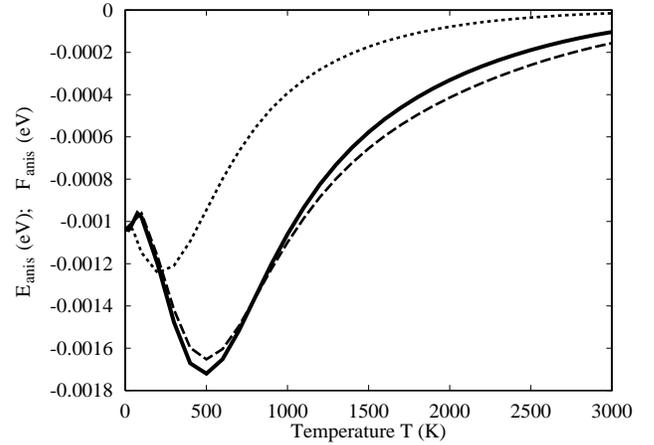,width=8.50cm}}
\vspace{.3cm}
\caption{
Temperature dependence of $F_{anis}(T)$ for a Fe-parametrized
$d$-band calculation for the monolayer with $d$-bandfilling $n_d=6$.  For the dashed curve,
only Fermi statistics is taken into account, for the solid curve the
lattice expansion is added, and the dotted curve includes the effects
of Fermi statistics, lattice expansion and entropy.}
\label{figtemp}
\end{figure}
\begin{figure}
\centerline{\psfig{figure=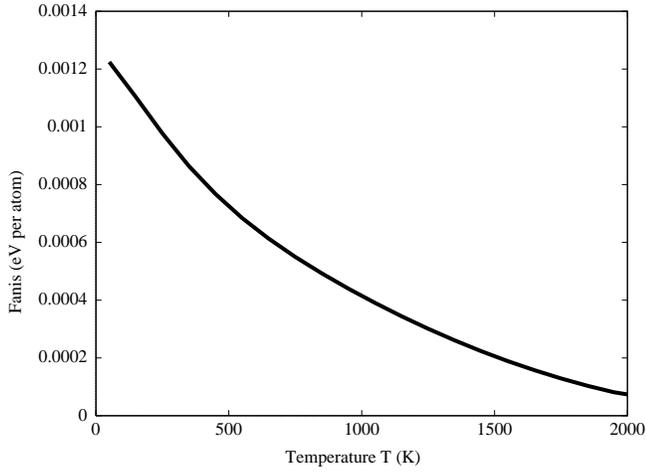,width=8.50cm}}
\vspace{.3cm}
\caption{
Temperature dependence of $F_{anis}(T)$ for a Fe-parametrized
$d$-band calculation  for 2 layers with $d$-bandfilling $n_d=6$.  The
calculation includes Fermi statistics, lattice expansion and entropy.}
\label{templ2}
\end{figure}
\begin{figure}
\centerline{\psfig{figure=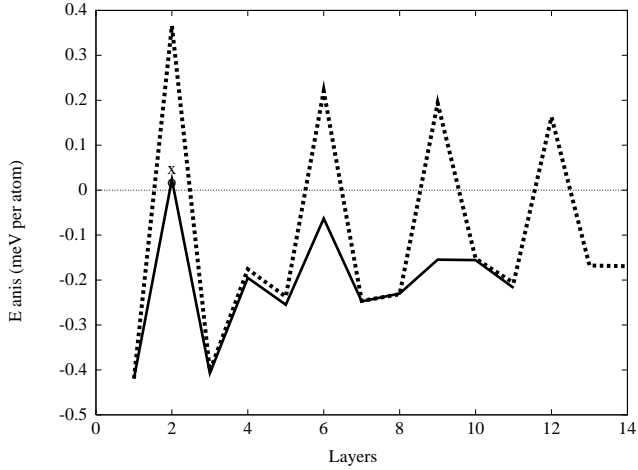,width=8.50cm}}
\vspace{.3cm}
\caption{
Magnetic anisotropy energy of Fe as a function of the number of layers 
calculated in the combined interpolation scheme.
The calculation for 1/4 BZ (dashed lines) yields periodic oscillations
caused by the incorrect symmetry of ${\bf E}_{anis}^{in-plane}$.
Summation over 1/2 BZ corrects this problem.    $\bullet$ and x are
calculations with 15356 and 108228 points in the 1/2 BZ respectively.}
\label{fe(l)}
\end{figure}

\begin{figure}
\vspace{2cm}
\centerline{\psfig{figure=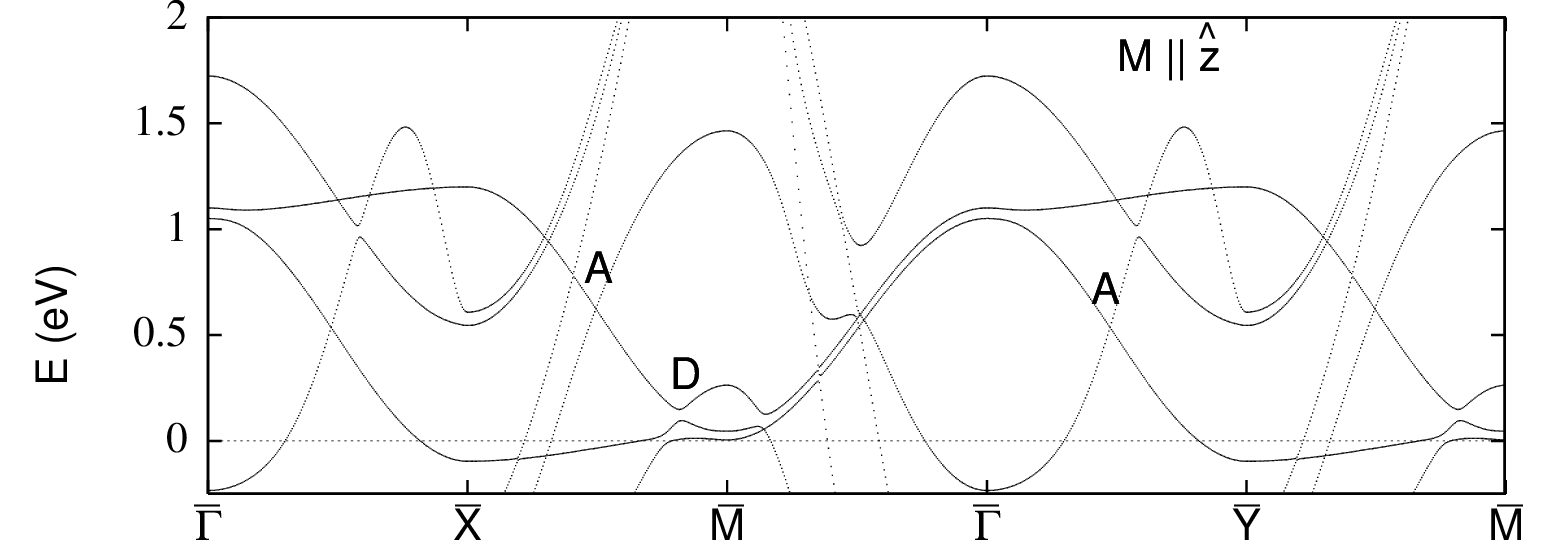,width=8.50cm,height=3.0cm}}
\vspace{.3cm}
\centerline{\psfig{figure=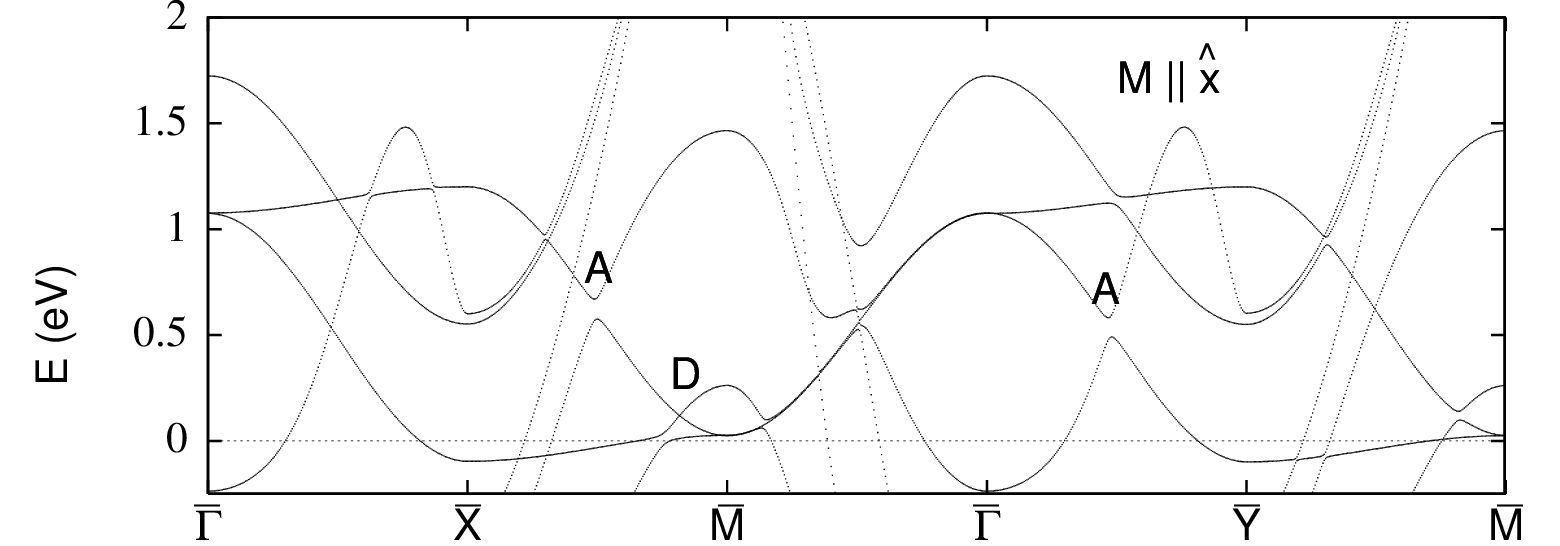,width=8.5cm,height=3.0cm}}
\vspace{.5cm}
\caption{Monolayer bandstructure of the $3s$- and $4s$-band for Fe-parameters, calculated within
  the combined interpolation scheme with the magnetization {\bf M}
  parallel to the layer normal $z$ in the upper part and in-plane parallel
  $x$ in the lower part.  High symmetry points are the same as in Fig.~\ref{figbandstr}.}
\label{band}
\end{figure}
\begin{figure}
\centerline{\psfig{figure=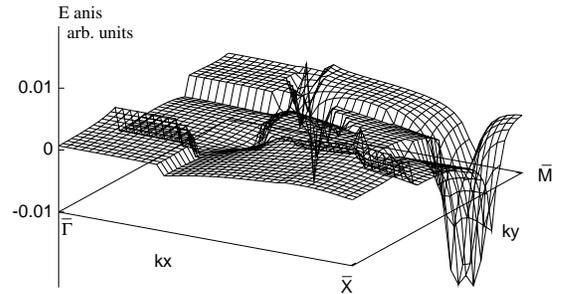,width=8.50cm}}
\vspace{2cm}
\caption{{\bf k}-space resolved anisotropy energy for the Fe monolayer at a $3s$- and
  $4d$-bandfilling of 7.8.  The ring-shaped dip near the M-point is caused
  by degeneracy D.}
\label{fe3d7_8}
\end{figure}
\begin{figure}
\centerline{\psfig{figure=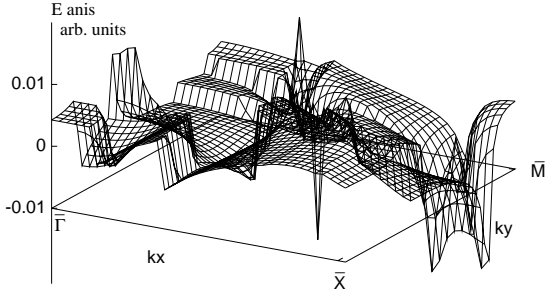,width=8.50cm}}
\vspace{.5cm}
\caption{{\bf k}-space resolved anisotropy energy of Fe for 3 layers at a $3s$- and
  $4d$-bandfilling of 8.0.  The negative peak near the M-point is caused
  by degeneracy D.}
\label{fe3dn3}
\vspace{1cm}
\end{figure}
\begin{figure}
\centerline{\psfig{figure=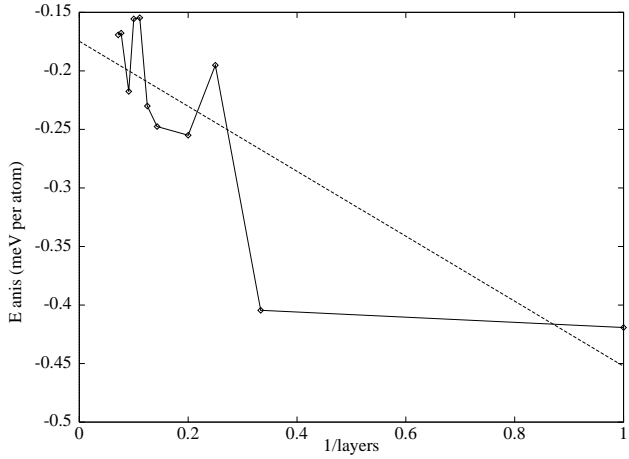,width=8.50cm}}
\vspace{.5cm}
\caption{Magnetic anisotropy energy of Fe as a function of the 1/$l$
  ($l$: number of layers).  Including the dipole-dipole anisotropy
  energy, we obtain in-plane magnetization from the fourth layer on.
  Linear least square fit yields
  \mbox{${\bf K}_v$ = -0.17 meV per atom} and \mbox{${\bf K}_s$ = -0.28 meV per atom}.}
\label{fe(1/l)}
\end{figure}
\begin{figure}
\centerline{\psfig{figure=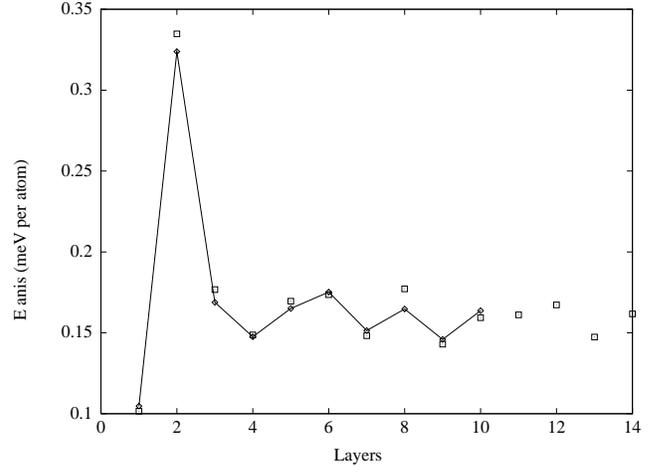,width=8.50cm}}
\vspace{.5cm}
\caption{
Magnetic anisotropy energy of Ni as a function of the number of
layers,  calculated within the combined interpolation scheme. $\Box$ is
a calculation for 1/4 BZ with 1722 points, the solid curve is a
calculation for 1/2 BZ with 6188 points.}
\label{Ni(l)}
\end{figure}

\end{twocolumn}

\begin{thebibliography}{99}
\bibitem {transfni} The parameters 
$A_i$ used by Weling and Callaway have to be transformed 
into the coordinate system used here by 
\[
\begin{array}{ll}
\tilde{B}_1={A}_4, & \quad \tilde{B}_4=-{A}_1, \\
\tilde{B}_2={A}_2-{A}_3, & \quad 
                   \tilde{B}_5=-\frac{1}{3}({A}_4+4{A}_5), \\
\tilde{B}_3={A}_2+{A}_3, & \quad
                   \tilde{B}_6=\frac{2}{\sqrt{3}}{A}_6
\end{array}
\]
and $J_{ex}=J_{ex}(e_g)$, $J'_{ex}=J_{ex}(t_{2g})$.
\bibitem {transffe} The corresponding
transformation formulas are
\[
\begin{array}{ll}
\tilde{B}_1=A_1, & \quad 
       \tilde{B}_4=\left(\frac{3}{4}A_4+\frac{1}{4}A_2\right), \\
\tilde{B}_2=A_2, & \quad 
       \tilde{B}_5=\left(\frac{1}{4}A_4+\frac{3}{4}A_2\right), \\
\tilde{B}_3=A_1, & \quad 
       \tilde{B}_6=\frac{\sqrt{3}}{4}(A_2-A_4)
\end{array}
\]
and $J'_{ex}=J_{ex}$.
\bibitem {angles} $\theta$ denotes the angle between the magnetization
  direction $z_M$ and the surface normal $\hat{z}$. $\phi$ is the angle
  between the $x$-axis and the projection of $z_M$ in the
  plane of the monolayer.
\end{thebibliography}
\end{document}